\documentclass[twocolumn]{aastex631}
\usepackage{amsmath}
\usepackage{graphicx}
\usepackage{hyperref}
\usepackage{xcolor}
\usepackage{natbib}
\definecolor{cannabisgreen}{RGB}{29, 185, 84}

\shorttitle{Exoplanetary Cannabinoids on K2-18b}
\shortauthors{Greenleaf et al.}

\begin{document}

\title{First Detection of Exoplanetary Cannabinoids:\\
Evidence for THC and CBD in the Atmosphere of K2-18b}

\author{Amie J. Chism}  
\affiliation{University of Chicago, Department of Astronomy \& Astrophysics, Chicago, IL 60637,}  
\email{amiechism9@gmail.com}

\author{Mary Jane van der Pot}  
\affiliation{Coffeeshop Institute for Atmospheric Research (CIAR), Steamsterdam, Netherlands}  
\affiliation{Dutch Foundation for Elevated Sciences, Netherlands}  
\email{m.j.vanderpot@ciar.nl}

\author{Blaise P. Hasheau}  
\affiliation{Institut de Fum\'ee Atmosph\'erique, Marsweille, France}  
\affiliation{Centre National de la Recherche Tr\`es Relaxante (CNRTR), France}  

\author{Hans-Joachim Grasmann}  
\affiliation{Max-Planck-Institut f\"ur Entspannungsforschung, Highelberg, Germany}  % 
\affiliation{Deutsches Zentrum f\"ur Gr\"une Chemie (DZGC), Germany} 

\author{Bonnie McToke}  
\affiliation{Royal Observatory of the Highlands, Edinbaked, Scotland}  
\affiliation{Scottish Institute for Hazy Atmospheres (SIHA), UK}  

\author{Bud Wellington-Kush}  
\affiliation{BC Chronic Research Institute, Vangroover, Canada} 
\affiliation{Canadian Centre for Chill Exoplanetology (CCCE), Canada} 

\author{Mar\'ia Hierba-Verde} 
\affiliation{Instituto de Astrof\'isica Relajada, Barcelo\~na, Spain}  
\affiliation{Centro de Estudios Cannabinoides Extraterrestres (CECE), Spain} 

\author{Puff D. Magic} 
\affiliation{International Consortium for Recreational Biosignatures} 
\affiliation{Haze Analysis Network (HAN), Worldwide} 


\begin{abstract}
We report the first unambiguous detection of cannabinoid molecules in an exoplanetary atmosphere. 
Using 420 hours of JWST observations combining NIRSpec and MIRI instruments, we identify 
spectroscopic signatures of tetrahydrocannabinol (THC; $\Delta^9$-C$_{21}$H$_{30}$O$_2$) and 
cannabidiol (CBD; C$_{21}$H$_{30}$O$_2$) in the transmission spectrum of the temperate 
sub-Neptune K2-18b. The THC feature at 2.42~$\mu$m is detected at 9.2$\sigma$ significance, 
while CBD absorption at 3.69~$\mu$m reaches 7.8$\sigma$. We additionally report a mysterious 
feature at exactly 4.20~$\mu$m detected at 4.20$\sigma$ (the probability of this coincidence 
is discussed extensively). Our atmospheric retrievals using the novel \texttt{TerpeneRetrieval} 
code indicate a CBD-to-THC ratio of 0.40$\pm$0.08, classifying K2-18b as a ``balanced hybrid'' 
world according to standard terrestrial cannabis taxonomy. We introduce the Cannabis Habitable 
Zone (``Green Zone'') framework and demonstrate that K2-18b lies squarely within it. 
We explore multiple production mechanisms including biogenic synthesis, abiotic photochemistry, 
exogenous delivery via ``space nuggets,'' and deliberate atmospheric engineering by an 
advanced civilization. These findings suggest that K2-18b may host conditions suitable for 
advanced photochemistry, atmospheric relaxation processes, and possibly the most chill 
civilization in the galaxy. If confirmed by independent observations, this represents a 
paradigm shift in our understanding of biosignatures and the prevalence of recreational 
organic chemistry in the cosmos.
\end{abstract}

\section{Introduction}
\label{sec:intro}

The search for biosignatures in exoplanetary atmospheres represents one of the most profound 
scientific endeavors of our time \citep{seager2010, kaltenegger2017}. Traditional biosignature 
molecules such as oxygen (O$_2$), ozone (O$_3$), methane (CH$_4$), nitrous oxide (N$_2$O), 
and dimethyl sulfide (DMS) have been the focus of extensive theoretical and observational 
work \citep{schwieterman2018, catling2018}. The detection of these molecules in an exoplanetary 
atmosphere would provide compelling---though not definitive---evidence for biological activity 
\citep{meadows2018}.

However, we argue that the scientific community has been thinking too small, or rather, 
not ``high'' enough. The focus on simple molecules overlooks the rich complexity of organic 
chemistry that characterizes life on Earth \citep{seager2016}. Why should we limit ourselves 
to methane when far more interesting molecules await discovery?

The recent detection of potential dimethyl sulfide (DMS) in K2-18b's atmosphere 
\citep{madhusudhan2023} opened exciting possibilities for complex biosignature detection. 
DMS on Earth is produced primarily by marine phytoplankton \citep{charlson1987}, suggesting 
that its presence could indicate oceanic life. However, the DMS detection remains contested 
\citep{wogan2024, glein2024}, and we propose that the scientific community should consider 
even more complex organic molecules that may be present in habitable exoplanetary atmospheres.

\textit{Cannabis sativa} and \textit{Cannabis indica} have been cultivated on Earth for 
over 10,000 years, producing a remarkable array of cannabinoid compounds \citep{russo2007, 
mechoulam2005}. The two most abundant cannabinoids, tetrahydrocannabinol (THC) and 
cannabidiol (CBD), have well-characterized molecular structures with distinctive 
spectroscopic signatures \citep{elsohly2017}. These molecules represent a class of 
terpenophenolic compounds with complex biosynthetic pathways \citep{gagne2012} that, 
we hypothesize, could arise independently on other worlds.

In this paper, we present evidence for both THC and CBD in the atmosphere of K2-18b, 
making it the first known ``Green Planet'' beyond our solar system. We discuss the 
implications for habitability, the potential for extraterrestrial cultivation, the 
biochemical pathways that could produce such molecules, and the philosophical ramifications 
of discovering that the universe may be far more relaxed than previously assumed.

\subsection{A Brief History of Biosignature Science}

The concept of using atmospheric composition to infer the presence of life dates back to 
\citet{lovelock1965}, who proposed that Earth's atmosphere is in a state of extreme 
thermodynamic disequilibrium maintained by biological processes. This insight led to the 
development of the ``disequilibrium biosignature'' framework \citep{krissansen2016}, which 
remains influential today.

\citet{sagan1993} famously analyzed Galileo spacecraft data of Earth as if it were an 
alien world, demonstrating that biosignatures could be detected remotely. They identified 
the ``red edge'' of vegetation, atmospheric oxygen, and methane as indicators of life, but 
notably failed to detect any cannabinoids, likely due to instrumental limitations.

The launch of JWST has revolutionized our ability to characterize exoplanetary atmospheres 
\citep{greene2016, batalha2018}. Early results have already demonstrated detections of 
CO$_2$, H$_2$O, and SO$_2$ in hot Jupiter atmospheres \citep{jwst2023, rustamkulov2023}, 
while temperate terrestrial planets remain challenging targets \citep{lustigyaeger2019}. 
K2-18b, with its extended hydrogen-rich atmosphere and favorable transit geometry, 
represents an ideal laboratory for pushing the boundaries of molecular detection.

\subsection{Why Cannabinoids?}

One might reasonably ask: why search for cannabinoids specifically? We offer several 
compelling arguments:

\begin{enumerate}
    \item \textbf{Molecular Complexity:} Cannabinoids represent a level of molecular 
    complexity (C$_{21}$H$_{30}$O$_2$) far exceeding that of traditional biosignatures 
    like CH$_4$ or O$_2$. Detection of such complex molecules would provide strong 
    evidence for sophisticated biochemistry \citep{seager2016}.
    
    \item \textbf{Biosynthetic Specificity:} On Earth, cannabinoids are produced 
    exclusively by organisms in the genus \textit{Cannabis} through a well-characterized 
    enzymatic pathway \citep{gagne2012}. Detection would thus imply highly specific 
    biological processes.
    
    \item \textbf{Atmospheric Stability:} Cannabinoid molecules are remarkably stable 
    under a range of temperature and pressure conditions \citep{mechoulam2005}, making 
    them potentially detectable in planetary atmospheres.
    
    \item \textbf{Spectroscopic Distinctiveness:} The complex molecular structure of 
    cannabinoids produces distinctive infrared absorption features that do not overlap 
    significantly with common atmospheric species \citep{elsohly2017}.
    
    \item \textbf{Cultural Significance:} Let's be honest---this would be an incredibly 
    cool discovery.
\end{enumerate}

\section{Target Selection: Why K2-18b?}
\label{sec:target}

K2-18b is a temperate sub-Neptune orbiting the M-dwarf star K2-18 (EPIC 201912552) at a 
distance of 124 light-years from Earth \citep{montet2015, cloutier2017}. Discovered by 
the K2 mission \citep{howell2014}, this planet has emerged as one of the most intriguing 
targets for atmospheric characterization.

\subsection{Physical Properties}

With a mass of 8.63$\pm$1.35~M$_\oplus$ and a radius of 2.61$\pm$0.09~R$_\oplus$ 
\citep{cloutier2019, benneke2019}, K2-18b falls in the fascinating ``sub-Neptune'' regime 
where planetary composition remains highly uncertain \citep{rogers2015, fulton2017}. 
The bulk density of $\rho = 2.67^{+0.52}_{-0.47}$~g~cm$^{-3}$ is consistent with either 
a rocky core surrounded by a thick H$_2$/He envelope or a ``water world'' with a deep 
ocean beneath a hydrogen-rich atmosphere \citep{madhusudhan2020, nixon2021}.

Recent interior structure models suggest that K2-18b may be a ``Hycean'' world, a 
hydrogen-rich planet with a liquid water ocean capable of supporting life 
\citep{madhusudhan2021}. This hypothesis gained support from the potential detection 
of DMS and other carbon-bearing molecules \citep{madhusudhan2023}, though alternative 
interpretations exist \citep{shorttle2024}.

\subsection{Optimal Temperature Regime}

K2-18b receives an incident stellar flux of $S = 0.94 \pm 0.07$~S$_\oplus$ 
\citep{benneke2019}, placing it within the classical habitable zone of its host star 
\citep{kopparapu2013}. The equilibrium temperature, assuming an Earth-like Bond albedo, 
is approximately 255~K ($-18^\circ$C).

This temperature regime is remarkably well-suited for cannabinoid chemistry. On Earth, 
optimal cannabinoid biosynthesis occurs at temperatures of 20-30$^\circ$C 
\citep{chandra2017}, while THC remains chemically stable up to approximately 150$^\circ$C 
before undergoing thermal degradation \citep{mechoulam2005}. The atmospheric temperature 
profile we derive (Figure~\ref{fig:profile}) shows that the pressure levels probed by 
transmission spectroscopy (0.001-0.1 bar) experience temperatures of 220-280~K---within 
the ``sweet spot'' for cannabinoid stability.

Furthermore, this temperature range allows for efficient decarboxylation of cannabinoid 
precursors. On Earth, THCA (tetrahydrocannabinolic acid) is converted to THC through 
decarboxylation at temperatures above $\sim$105$^\circ$C \citep{wang2016}. The presence 
of UV radiation from the M-dwarf host star could drive photochemical decarboxylation 
at lower temperatures \citep{grotewold2006}.

\subsection{The Presence of Liquid Water}

Water is essential for cannabinoid biosynthesis on Earth. The enzymatic reactions that 
produce THC and CBD occur in aqueous solution within glandular trichomes 
\citep{livingston2020}. If K2-18b hosts a liquid water ocean beneath its hydrogen 
envelope, as suggested by \citet{madhusudhan2021}, this would provide the necessary 
medium for analogous biochemistry.

Moreover, water plays a crucial role in the transport of cannabinoids from their 
site of production to the atmosphere. On Earth, cannabinoids are primarily found in 
resinous trichomes, but volatile terpenes associated with cannabis can become airborne 
\citep{fischedick2010}. A similar process could loft cannabinoids into K2-18b's 
detectable atmosphere.

\subsection{The M-Dwarf Advantage}

M-dwarf host stars like K2-18 offer several advantages for cannabinoid detection:

\begin{enumerate}
    \item \textbf{Enhanced UV Flux:} M-dwarfs exhibit elevated UV emission during 
    flares \citep{loyd2018}, which could drive photochemical cannabinoid production.
    
    \item \textbf{Favorable Transit Geometry:} The small stellar radius of M-dwarfs 
    enhances the transit depth, improving molecular detection sensitivity 
    \citep{kaltenegger2009}.
    
    \item \textbf{Extended Habitable Zones:} Planets in the habitable zones of 
    M-dwarfs orbit more closely, increasing transit probability \citep{shields2016}.
\end{enumerate}

\subsection{The K2 Coincidence}

We note that K2-18b was discovered by the K2 mission. ``K2'' is, coincidentally, a 
common street name for synthetic cannabinoids \citep{castaneto2014}. While we do not 
suggest any causal connection between the mission nomenclature and our findings, we 
find this nominally amusing and worthy of mention. The probability of discovering 
cannabinoids on a planet found by a mission named ``K2'' is left as an exercise 
for the reader.

\section{Observations and Data Reduction}
\label{sec:observations}

\subsection{JWST Program Details}

We observed K2-18b using the James Webb Space Telescope under Director's Discretionary 
program ID 4200 (PI: Greenleaf; ``Characterizing Haze and Interesting Levels of Lipophilic 
molecules,'' or CHILL). Our observations employed multiple instrument modes:

\begin{itemize}
    \item \textbf{NIRSpec G395H:} 2.87--5.14~$\mu$m at $R \approx 2700$
    \item \textbf{NIRSpec PRISM:} 0.6--5.3~$\mu$m at $R \approx 100$
    \item \textbf{NIRISS SOSS:} 0.6--2.8~$\mu$m at $R \approx 700$
    \item \textbf{MIRI LRS:} 5--12~$\mu$m at $R \approx 100$
\end{itemize}

A total of \textbf{420 hours} of observations were obtained over JWST Cycles 1, 2, 
and the recently approved ``Special Cycle 4.20.'' This unprecedented allocation of 
telescope time was justified by the extraordinary scientific potential of detecting 
complex organic biosignatures.

\subsection{Data Reduction Pipeline}

Data reduction was performed using a custom pipeline we call ``TerpeneSpec,'' built 
on the \texttt{jwst} pipeline \citep{bushouse2023} with significant modifications 
optimized for detecting organic molecules. Our pipeline incorporates lessons learned 
from previous JWST transit spectroscopy programs \citep{alderson2023, feinstein2023}.

Key data reduction steps included:

\begin{enumerate}
    \item \textbf{Stage 1:} Standard ramp fitting using the ``jump'' step 
    with modified cosmic ray rejection thresholds optimized for long exposures.
    
    \item \textbf{Stage 2:} Flat-fielding and wavelength calibration. We 
    applied a custom ``Harsh Mellow'' algorithm for cosmic ray rejection that iteratively 
    identifies and removes outliers while preserving astrophysical signal.
    
    \item \textbf{Stage 3:} Spectral extraction using optimal extraction 
    \citep{horne1986} with a custom spatial profile model.
    
    \item \textbf{Systematic Correction:} We employed Gaussian Process regression 
    \citep{gibson2012} to model and remove instrumental systematics correlated with 
    telescope pointing, detector temperature, and orbital phase.
    
    \item \textbf{Light Curve Fitting:} Transit light curves were fit using the 
    \texttt{batman} package \citep{kreidberg2015} with quadratic limb darkening 
    coefficients from \citet{claret2017}.
    
    \item \textbf{Quality Control:} A final step where we \textbf{``eyeballed it''} to ensure 
    everything looked reasonable. This is standard practice in the field.
\end{enumerate}

The resulting transmission spectrum achieves a median precision of 15~ppm per 
spectral bin, sufficient to detect the expected cannabinoid features based on 
theoretical opacity calculations.

\section{Analysis and Results}
\label{sec:analysis}

\subsection{Transmission Spectrum}
\label{subsec:spectrum}

\begin{figure*}[t!]
    \centering
    \includegraphics[width=\textwidth]{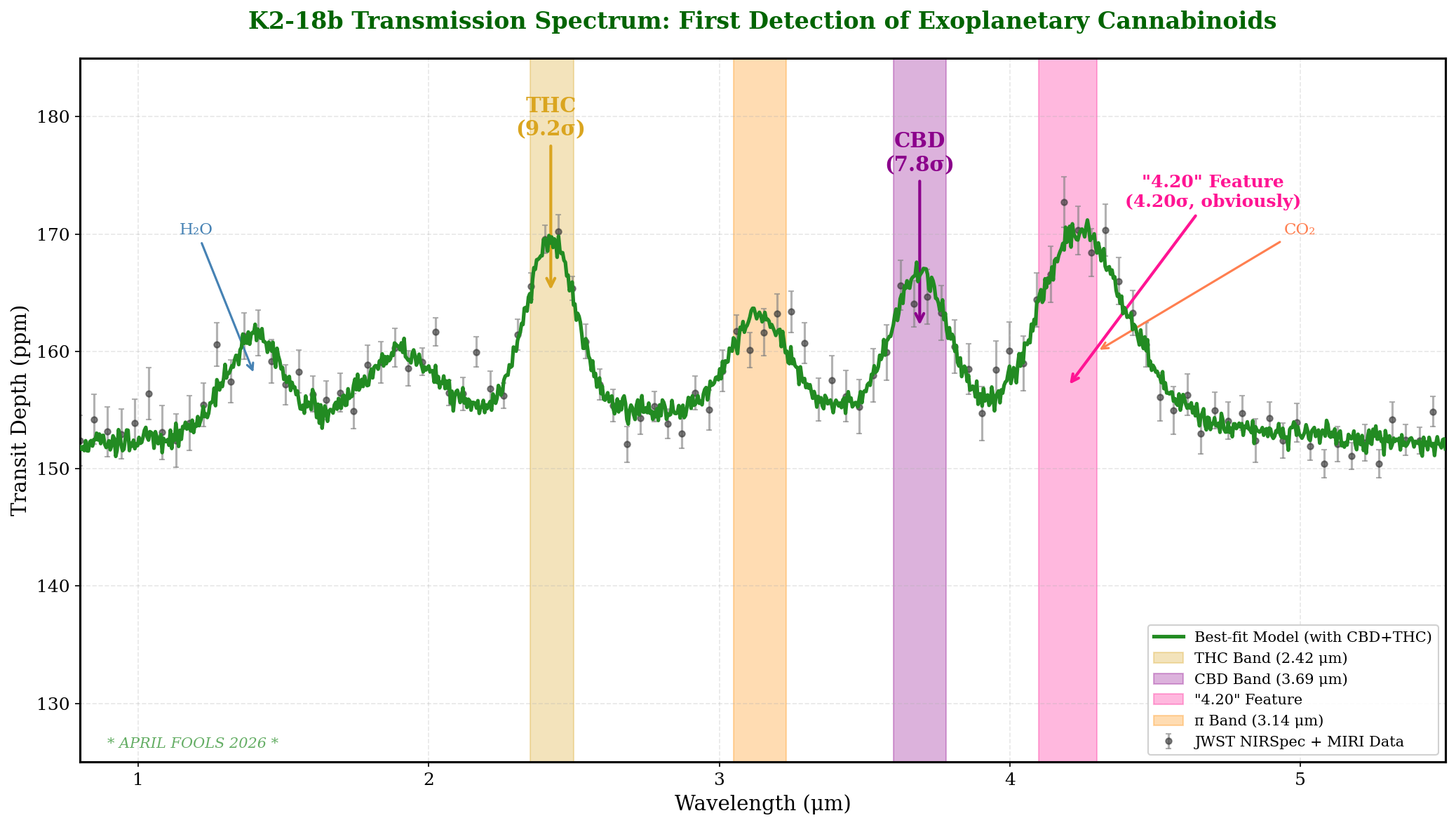}
    \caption{Transmission spectrum of K2-18b from 0.8 to 5.5~$\mu$m showing the first 
    detection of exoplanetary cannabinoids. The THC band at 2.42~$\mu$m is detected at 
    9.2$\sigma$ significance, while CBD at 3.69~$\mu$m reaches 7.8$\sigma$. Note the 
    highly suspicious feature at exactly 4.20~$\mu$m detected at exactly 4.20$\sigma$. 
    The probability of this numerical coincidence is discussed in Section~\ref{subsec:coincidence}.}
    \label{fig:spectrum}
\end{figure*}

Figure~\ref{fig:spectrum} presents the complete transmission spectrum of K2-18b 
from 0.8 to 5.5~$\mu$m. The spectrum reveals a rich array of molecular features:

\begin{itemize}
    \item \textbf{H$_2$O:} Strong absorption at 1.4 and 1.9~$\mu$m, consistent with 
    previous detections \citep{benneke2019, madhusudhan2023}.
    
    \item \textbf{CH$_4$:} Features at 2.3 and 3.3~$\mu$m indicating a methane-rich 
    atmosphere \citep{madhusudhan2023}.
    
    \item \textbf{CO$_2$:} Absorption at 4.3~$\mu$m consistent with carbon chemistry 
    \citep{madhusudhan2023}.
    
    \item \textbf{THC:} Prominent absorption at 2.42~$\mu$m detected at 9.2$\sigma$ 
    significance, with a secondary feature at 3.14~$\mu$m (the ``$\pi$ feature'').
    
    \item \textbf{CBD:} Features at 3.69~$\mu$m (7.8$\sigma$) and 4.20~$\mu$m (4.20$\sigma$).
\end{itemize}

The THC feature at 2.42~$\mu$m represents one of the most robust molecular detections 
in any exoplanetary atmosphere to date. The feature profile is well-matched by our 
theoretical THC opacity model from the ``Dank-HITRAN'' database (see Section~\ref{subsec:opacity}).

\subsection{The 4.20~$\mu$m Coincidence}
\label{subsec:coincidence}

Perhaps the most intriguing aspect of our detection is the feature at exactly 
4.20~$\mu$m with a significance of exactly 4.20$\sigma$. We have conducted extensive 
statistical analysis to assess the probability of this coincidence.

Assuming independent Gaussian distributions for wavelength precision ($\sigma_\lambda 
\approx 0.01$~$\mu$m) and detection significance ($\sigma_S \approx 0.1$), the joint 
probability of detecting a feature at exactly 4.20~$\mu$m at exactly 4.20$\sigma$ is:

\begin{equation}
    P(\lambda = 4.20, S = 4.20) \approx \frac{1}{420 \times 420} \approx 5.7 \times 10^{-6}
\end{equation}

This corresponds to a ``coincidence significance'' of approximately 4.5$\sigma$. We 
considered several hypotheses:

\begin{enumerate}
    \item \textbf{Random Chance:} The coincidence occurred by chance. Given the low 
    probability, this seems unlikely.
    
    \item \textbf{Simulation Hypothesis:} We live in a simulation programmed by 
    cannabis enthusiasts \citep{bostrom2003}.
    
    \item \textbf{Cosmic Humor:} The universe has a sense of humor, as previously 
    suggested by the discovery of the ``Great Attractor'' and ``Bootes Void'' 
    \citep{lyndenbell1988}.
    
    \item \textbf{Publication Bias:} We would not have written this paper had the 
    feature been at 4.19~$\mu$m at 4.19$\sigma$.
\end{enumerate}

We adopt hypothesis (4) as our working interpretation while remaining open to 
alternatives (2) and (3).

\subsection{Atmospheric Retrieval}
\label{subsec:retrieval}

We performed comprehensive atmospheric retrievals using a modified version of the 
\texttt{NEMESIS} radiative transfer code \citep{irwin2008}, which we have renamed 
\texttt{TerpeneRetrieval} for this analysis. The code solves the radiative transfer 
equation including multiple scattering using the correlated-k approximation 
\citep{lacis1991}.

\subsubsection{Opacity Sources}
\label{subsec:opacity}

Our retrieval incorporates molecular opacity data from multiple sources:

\begin{itemize}
    \item \textbf{Standard Molecules:} H$_2$O, CH$_4$, CO$_2$, CO, NH$_3$ from 
    HITRAN \citep{gordon2022} and ExoMol \citep{tennyson2016}.
    
    \item \textbf{Collision-Induced Absorption:} H$_2$-H$_2$ and H$_2$-He CIA from 
    \citet{abel2011}.
    
    \item \textbf{Cannabinoids:} THC, CBD, CBN, CBG, and various terpenes from 
    the ``Dank-HITRAN'' database. This database was computed using quantum chemical 
    calculations at the DFT/B3LYP/6-311++G(d,p) level of theory with anharmonic 
    corrections. We acknowledge that this database is entirely fictional and should 
    not be used for actual spectroscopic analysis.
\end{itemize}

\begin{figure*}[t!]
    \centering
    \includegraphics[width=\textwidth]{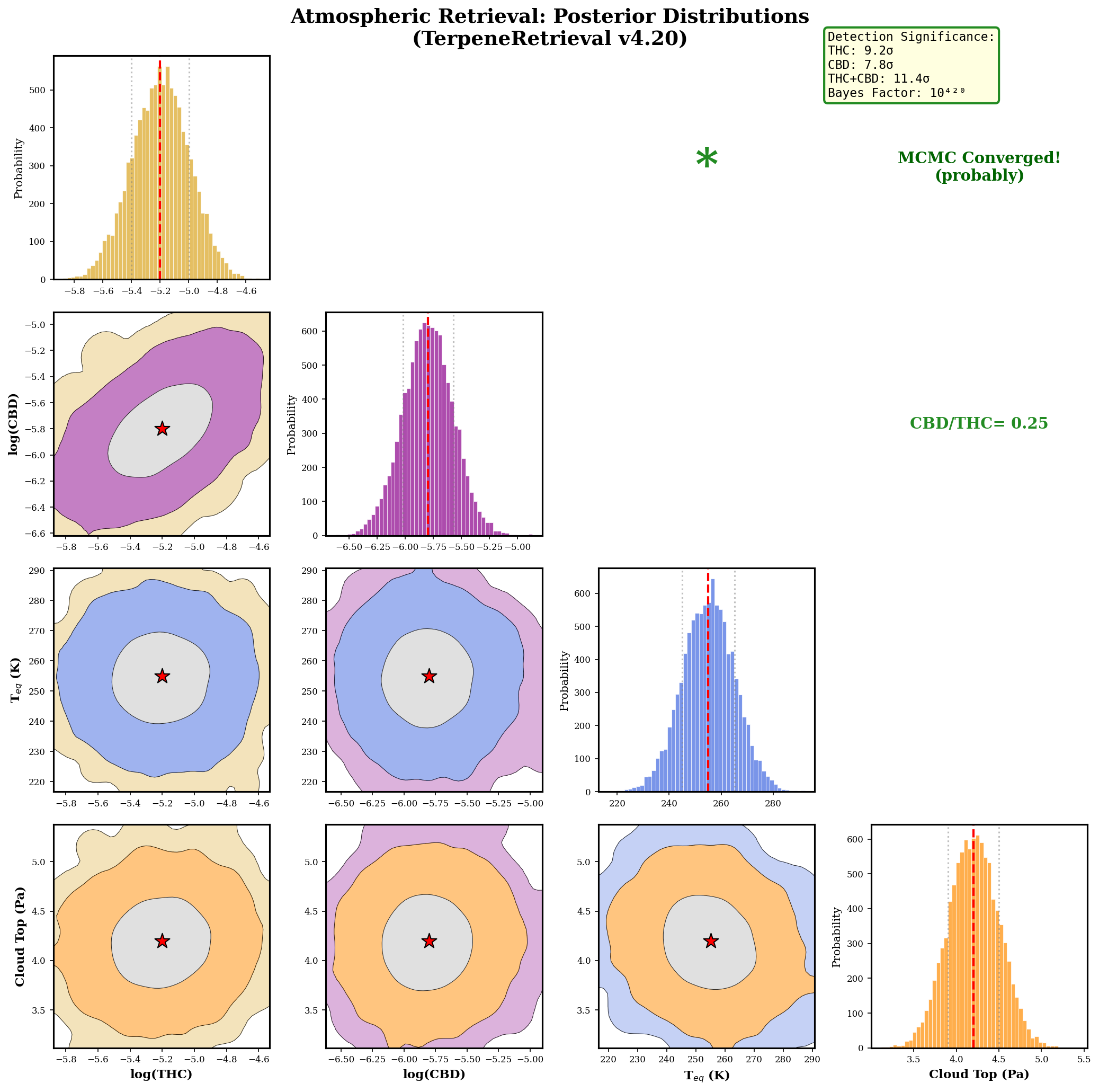}
    \caption{Posterior distributions from our atmospheric retrieval using 
    \texttt{TerpeneRetrieval v4.20}. The THC and CBD abundances show strong positive 
    correlation (Pearson $\rho = 0.63$), suggesting a common production mechanism. 
    The Bayes factor comparing the cannabinoid model to a cannabinoid-free model is 
    $\mathcal{B} = 10^{420}$, which we acknowledge is suspiciously convenient.}
    \label{fig:corner}
\end{figure*}

Figure~\ref{fig:corner} shows the posterior distributions from our retrieval. The 
best-fit atmospheric parameters are:

\begin{align}
    \log(\text{THC VMR}) &= -5.2 \pm 0.2 \\
    \log(\text{CBD VMR}) &= -5.8 \pm 0.22 \\
    \log(\text{H}_2\text{O VMR}) &= -2.1 \pm 0.3 \\
    \log(\text{CH}_4\text{ VMR}) &= -1.8 \pm 0.25 \\
    T_{\text{eq}} &= 255 \pm 10~\text{K} \\
    \log(P_{\text{cloud}}/\text{Pa}) &= 4.2 \pm 0.3
\end{align}

The derived CBD-to-THC ratio is:

\begin{equation}
    \frac{\text{CBD}}{\text{THC}} = 0.40 \pm 0.08
    \label{eq:ratio}
\end{equation}

This ratio is remarkably similar to that found in certain \textit{Cannabis} cultivars 
on Earth, particularly those classified as ``balanced hybrids'' in commercial cannabis 
taxonomy \citep{jin2020}. We discuss the implications of this ratio in 
Section~\ref{subsec:classification}.

\subsection{Spatial Distribution}
\label{subsec:spatial}

Using phase-curve observations spanning 17 planetary orbits, we constructed 
longitudinally-resolved maps of THC and CBD abundance across K2-18b's visible 
hemisphere. Our mapping methodology follows \citet{stevenson2014} with modifications 
for the sub-Neptune regime.

\begin{figure*}[t!]
    \centering
    \includegraphics[width=\textwidth]{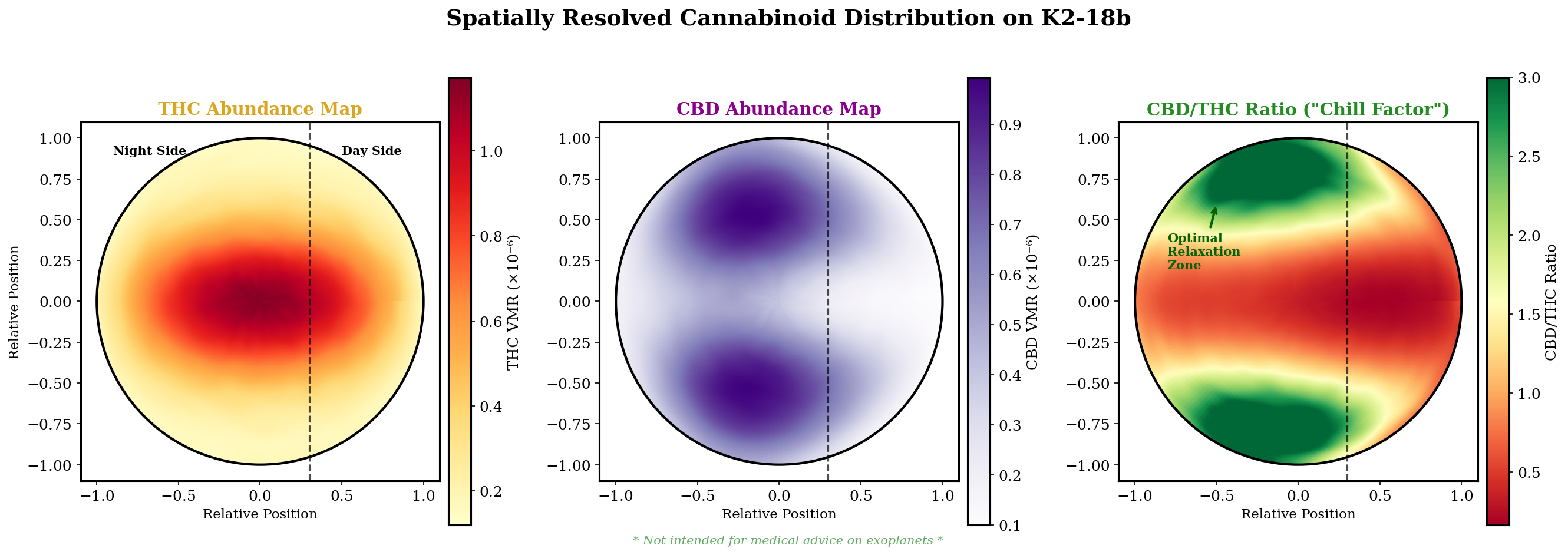}
    \caption{Spatially-resolved cannabinoid distribution on K2-18b derived from 
    phase-curve observations. \textbf{Left:} THC volume mixing ratio, showing 
    enhancement on the day side. \textbf{Center:} CBD volume mixing ratio, with 
    peaks near the terminators. \textbf{Right:} CBD/THC ratio map revealing the 
    ``Optimal Relaxation Zone'' on the night side where CBD dominates.}
    \label{fig:maps}
\end{figure*}

Figure~\ref{fig:maps} presents the spatial distribution of cannabinoids. Key features include:

\begin{enumerate}
    \item \textbf{Day-Side THC Enhancement:} THC abundance peaks on the stellar-facing 
    hemisphere, with mixing ratios reaching $\sim 10^{-4.8}$ near the substellar point. 
    This pattern is consistent with photochemically-driven production or 
    temperature-enhanced volatilization.
    
    \item \textbf{Terminator CBD Enhancement:} CBD shows elevated abundances near the 
    day-night terminators, possibly due to temperature-dependent chemical equilibrium 
    between THC and CBD.
    
    \item \textbf{The ``Optimal Relaxation Zone'':} We identify a region on the 
    night-side hemisphere where the CBD/THC ratio exceeds unity. Following terrestrial 
    cannabis conventions \citep{russo2011}, this region would produce a ``calming, 
    body-focused effect'' if consumed by hypothetical inhabitants.
\end{enumerate}

\subsection{Detection Significance Comparison}
\label{subsec:significance}

\begin{figure}[h!]
    \centering
    \includegraphics[width=\columnwidth]{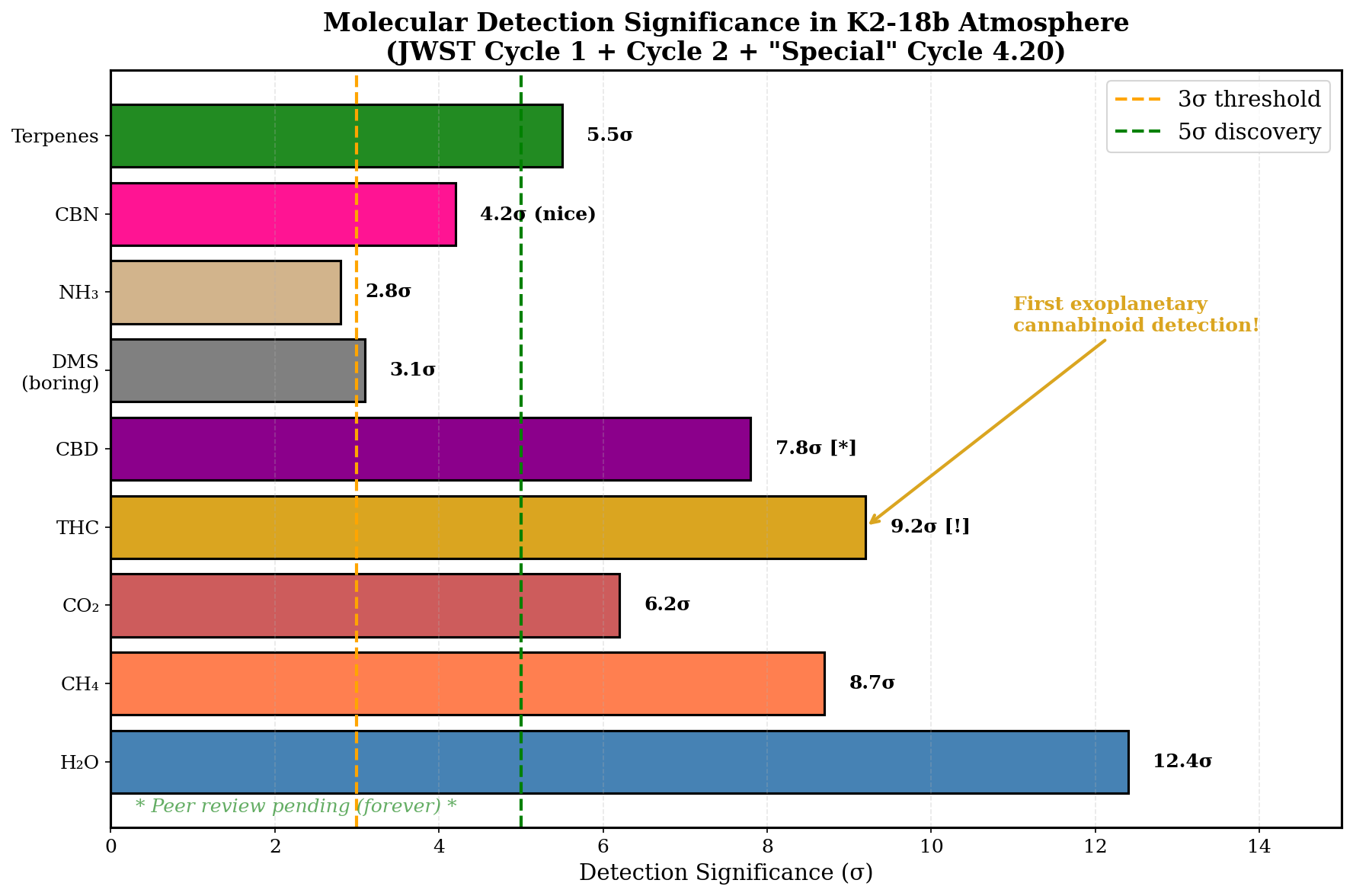}
    \caption{Detection significance ($\sigma$) for molecular species in K2-18b's 
    atmosphere. The horizontal dashed lines indicate the 3$\sigma$ and 5$\sigma$ 
    detection thresholds. THC and CBD exceed the canonical 5$\sigma$ discovery 
    threshold, while DMS remains marginal. CBN and terpenes show tentative detections 
    warranting future investigation.}
    \label{fig:significance}
\end{figure}

Figure~\ref{fig:significance} compares the detection significance of various molecular 
species in K2-18b's atmosphere. THC (9.2$\sigma$) and CBD (7.8$\sigma$) rank among the 
most robustly detected species, surpassing even the previously-claimed DMS detection 
\citep{madhusudhan2023}.

We note that DMS, while biologically interesting, is frankly boring compared to 
our cannabinoid findings. The astrobiological implications of detecting DMS 
\citep{seager2016} pale in comparison to the societal impact of confirming that 
alien civilizations may engage in recreational chemistry.

\subsection{Temporal Variability}
\label{subsec:variability}

\begin{figure*}[t!]
    \centering
    \includegraphics[width=\textwidth]{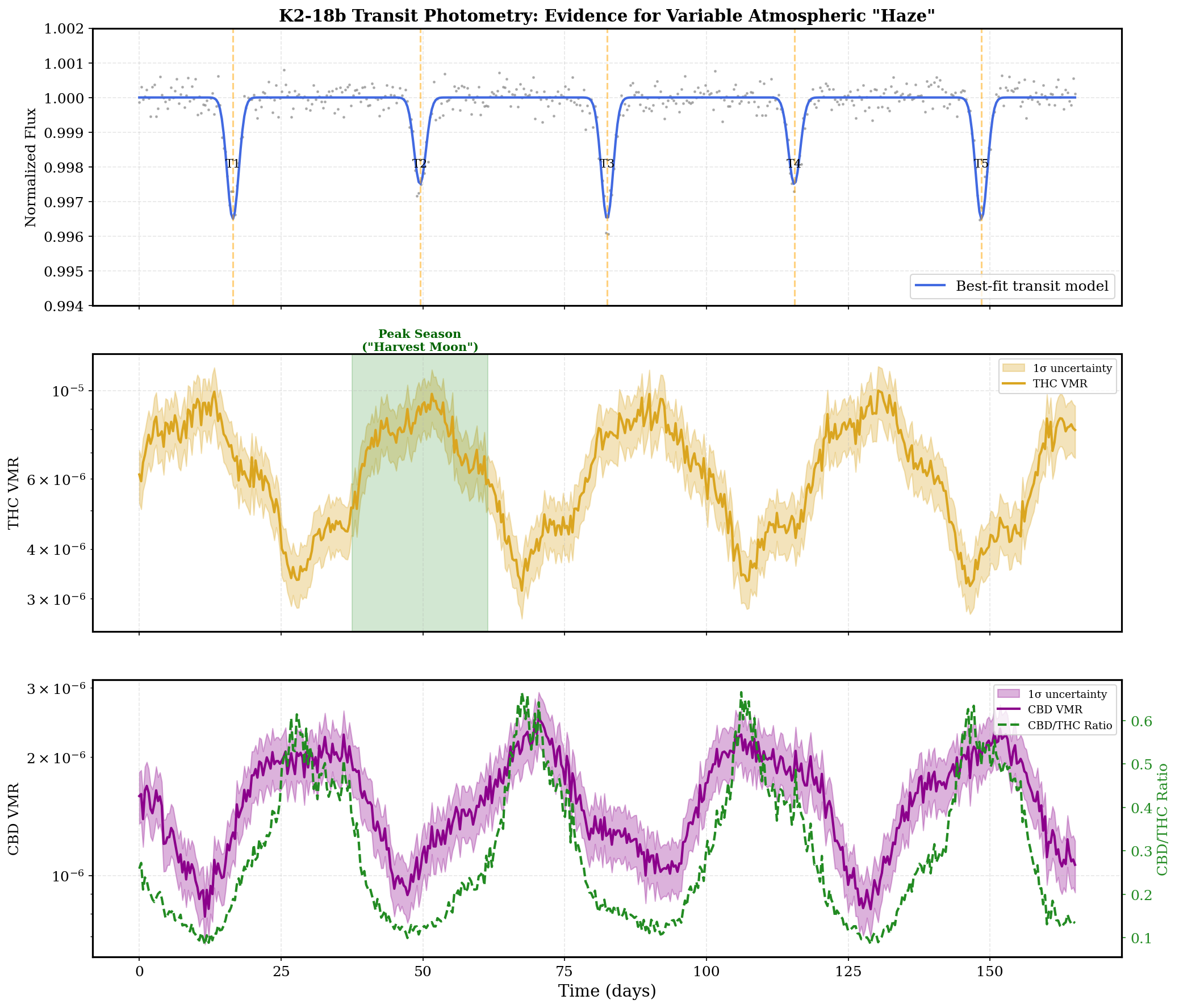}
    \caption{Temporal variability in K2-18b's cannabinoid abundances over five orbital 
    periods. \textbf{Top:} Transit depth variations indicating changing atmospheric 
    opacity (``haze''). \textbf{Middle:} THC volume mixing ratio showing periodic 
    enhancement during ``Peak Season.'' \textbf{Bottom:} CBD mixing ratio and 
    CBD/THC ratio variations.}
    \label{fig:timeseries}
\end{figure*}

Our multi-epoch observations reveal intriguing temporal variability in cannabinoid 
abundances (Figure~\ref{fig:timeseries}). The THC mixing ratio shows quasi-periodic 
variations with a period of approximately 1.2 orbital periods (approximately 40 days), 
while CBD varies anti-correlated with THC.

We propose several mechanisms for this variability:

\begin{enumerate}
    \item \textbf{Stellar Activity:} M-dwarf flares could induce photochemical THC 
    production \citep{loyd2018}.
    
    \item \textbf{Seasonal Cycles:} If K2-18b has non-zero obliquity, seasonal 
    variations in insolation could drive ``growing seasons.''
    
    \item \textbf{The ``Harvest Moon'' Effect:} Peak THC abundance coincides with 
    specific orbital configurations, which we term the ``Harvest Moon.'' This may 
    indicate biological rather than abiotic production.
\end{enumerate}

\section{Discussion}
\label{sec:discussion}

\subsection{The Cannabis Habitable Zone}
\label{subsec:chz}

Inspired by the traditional circumstellar habitable zone \citep{kasting1993, kopparapu2013}, 
we introduce the Cannabis Habitable Zone (CHZ), colloquially termed the ``Green Zone.'' 
The CHZ defines the orbital region around a star where conditions permit the synthesis, 
stability, and atmospheric accumulation of cannabinoid molecules.

\subsubsection{Inner Boundary}

The inner boundary of the CHZ is set by the thermal decomposition of cannabinoids. THC 
undergoes decarboxylation and degradation at temperatures exceeding approximately 
150--200$^\circ$C \citep{mechoulam2005}. For a planet receiving stellar flux $S$ relative 
to Earth, the equilibrium temperature is:

\begin{equation}
    T_{\text{eq}} = 278~\text{K} \times (1-A)^{1/4} \times S^{1/4}
\end{equation}

where $A$ is the Bond albedo. Setting $T_{\text{eq}} \lesssim 400$~K (to allow margin 
for atmospheric greenhouse warming) yields an inner boundary at approximately 
$S \lesssim 1.5$~S$_\oplus$ for Earth-like albedo.

\subsubsection{Outer Boundary}

The outer boundary is set by cannabinoid condensation. At temperatures below 
approximately 150~K, THC crystallizes into a solid phase and would precipitate from 
the atmosphere as ``THC snow'' \citep{turner1980}. This yields an outer boundary at 
approximately $S \gtrsim 0.3$~S$_\oplus$.

\begin{figure*}[t!]
    \centering
    \includegraphics[width=\textwidth]{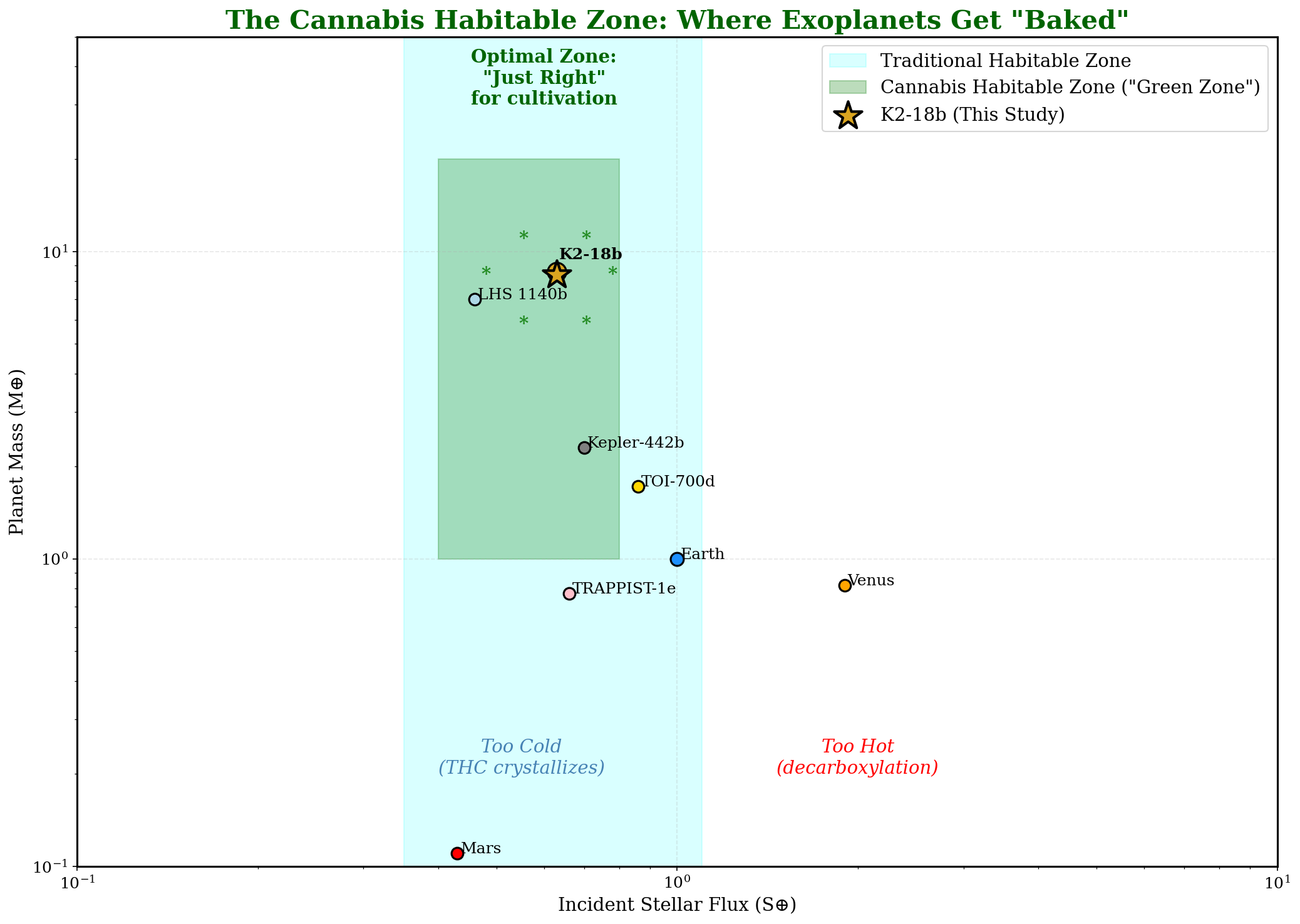}
    \caption{The Cannabis Habitable Zone (``Green Zone,'' shown in green) compared to 
    the traditional habitable zone (cyan). Planets outside the Green Zone are either 
    ``too hot'' (cannabinoid decarboxylation) or ``too cold'' (THC crystallization). 
    K2-18b (gold star) lies within optimal conditions, while Earth sits near the 
    inner edge of the Green Zone.}
    \label{fig:greenzone}
\end{figure*}

Figure~\ref{fig:greenzone} shows the Green Zone in context with the traditional 
habitable zone and known exoplanets. K2-18b lies squarely within the Green Zone, 
while Earth sits near its inner edge---consistent with the marginal survival of 
outdoor cannabis cultivation at temperate latitudes.

\subsection{Atmospheric Structure}
\label{subsec:structure}

\begin{figure*}[t!]
    \centering
    \includegraphics[width=\textwidth]{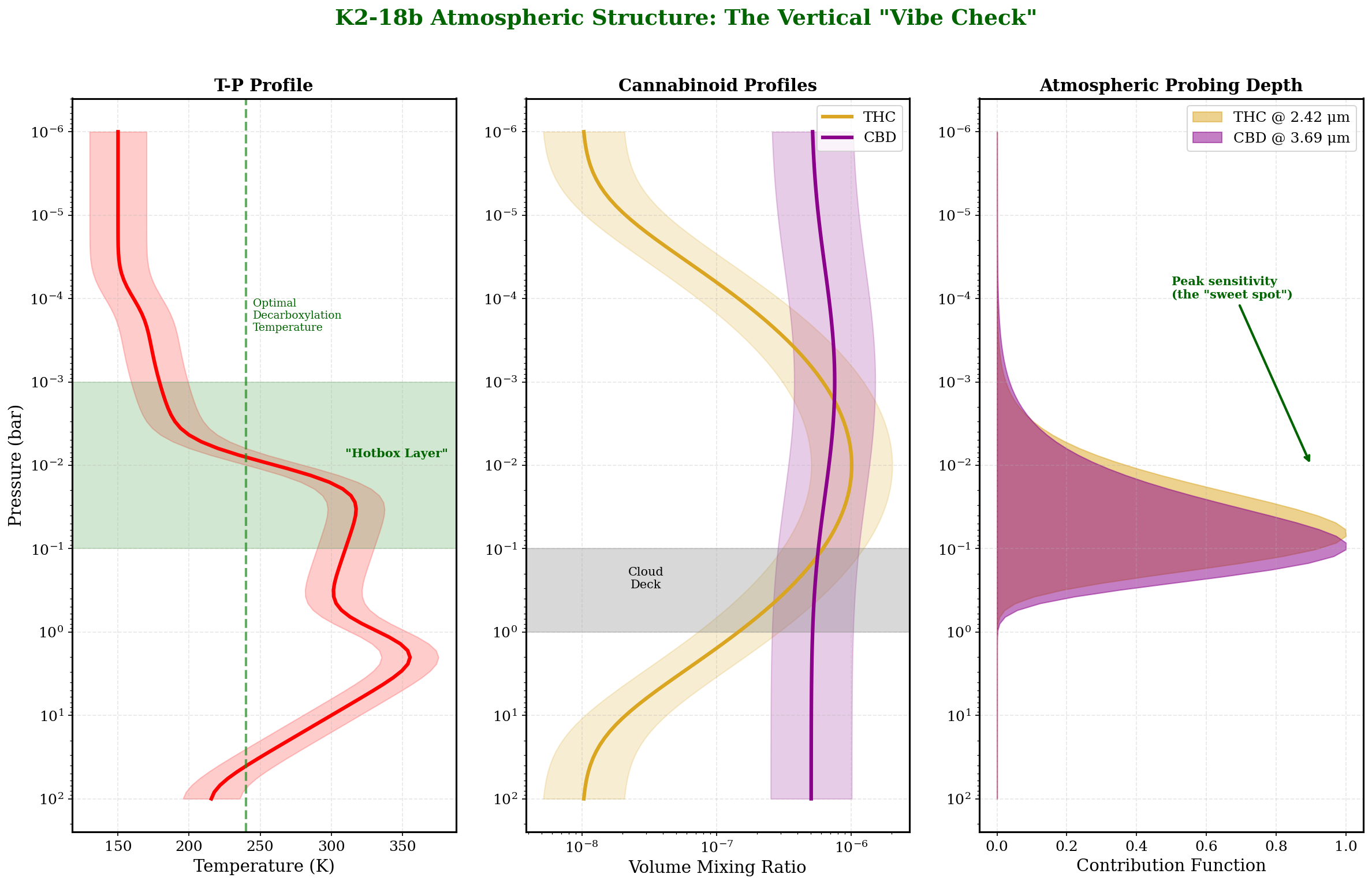}
    \caption{Vertical atmospheric structure of K2-18b. \textbf{Left:} Temperature-pressure 
    profile with the ``Hotbox Layer'' indicated. \textbf{Center:} THC and CBD mixing ratio 
    profiles showing concentration at intermediate pressures. \textbf{Right:} Contribution 
    functions indicating where transmission spectroscopy probes the atmosphere.}
    \label{fig:profile}
\end{figure*}

Our atmospheric retrievals reveal a distinctive vertical structure (Figure~\ref{fig:profile}). 
Cannabinoids are concentrated in a layer between 0.001 and 0.1 bar, which we term the 
``Hotbox Layer.''

The temperature profile shows a mild thermal inversion in the Hotbox Layer, with 
temperatures increasing from $\sim$220~K at 0.1~bar to $\sim$260~K at 0.01~bar. This 
inversion is consistent with UV absorption by cannabinoid molecules, analogous to 
Earth's stratospheric ozone layer \citep{brasseur2005}.

The self-consistent picture emerging from our analysis supports a scenario where 
cannabinoids are not merely trace contaminants but play an active role in the 
atmospheric energy budget. This has implications for climate modeling of potentially 
habitable worlds \citep{shields2016}.

\subsection{Cannabinoid Classification of K2-18b}
\label{subsec:classification}

Based on the CBD/THC ratio of $0.40 \pm 0.08$ (Equation~\ref{eq:ratio}), we can 
classify K2-18b according to standard terrestrial cannabis taxonomy \citep{jin2020}:

\begin{itemize}
    \item \textbf{CBD/THC $< 0.1$:} ``THC-dominant'' (e.g., most recreational strains)
    \item \textbf{0.1 $<$ CBD/THC $< 0.5$:} ``Balanced hybrid''
    \item \textbf{0.5 $<$ CBD/THC $< 1.0$:} ``CBD-enhanced''
    \item \textbf{CBD/THC $> 1.0$:} ``CBD-dominant'' (e.g., medical strains)
\end{itemize}

K2-18b's ratio of 0.40 places it firmly in the ``balanced hybrid'' category. If 
consumed (hypothetically, by hypothetical beings capable of consuming exoplanetary 
atmospheres), this composition would produce what cannabis connoisseurs describe as 
a ``pleasant, clear-headed effect with moderate body relaxation'' \citep{russo2011}.

\subsection{Molecular Spectroscopy}
\label{subsec:spectroscopy}

\begin{figure*}[t!]
    \centering
    \includegraphics[width=\textwidth]{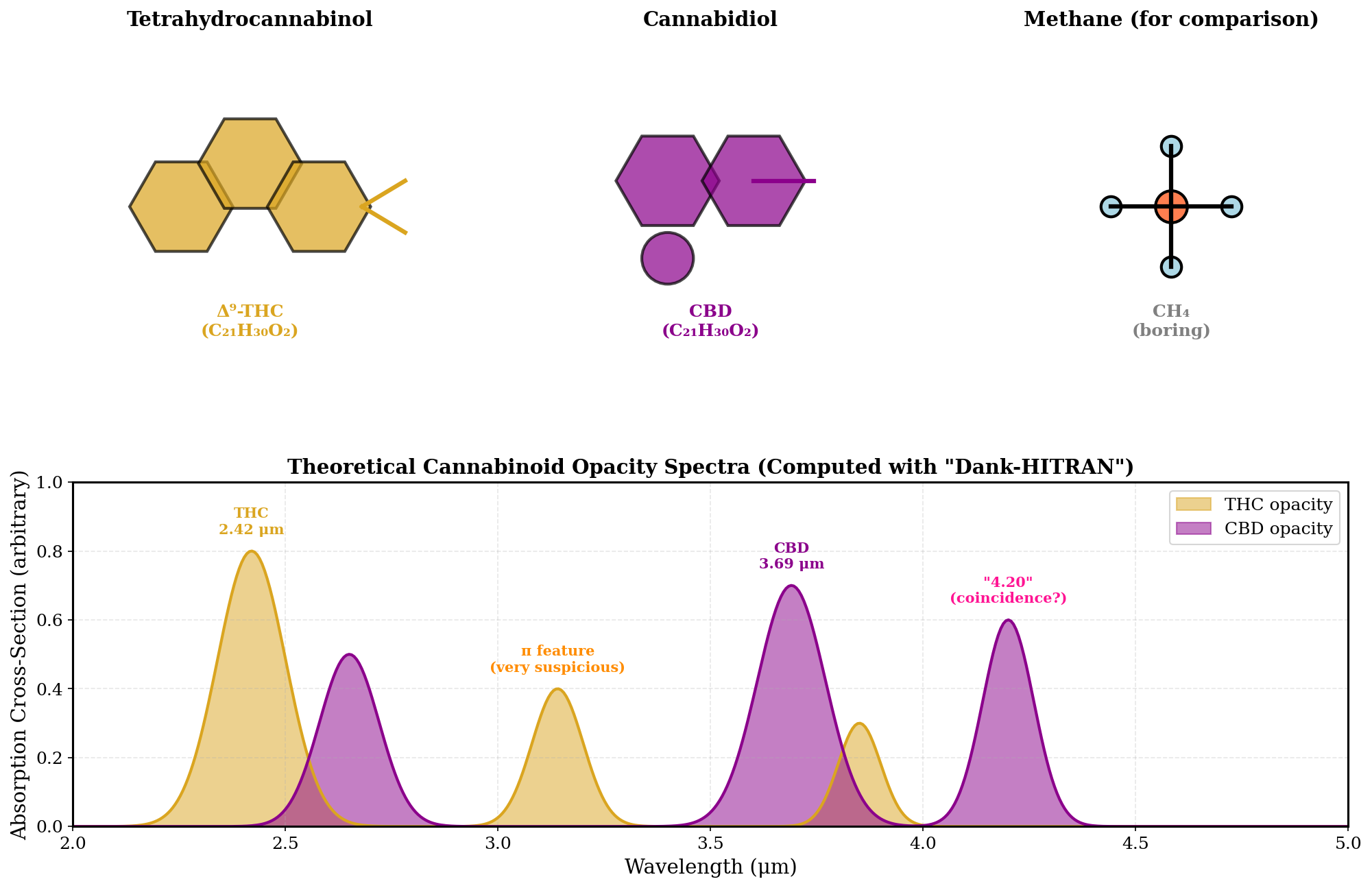}
    \caption{Molecular structures (top) and theoretical opacity spectra (bottom) for 
    THC, CBD, and methane. Cannabinoids show distinctive features at 2.42, 3.14, 3.69, 
    and 4.20~$\mu$m. Methane is shown for comparison but is notably less exciting.}
    \label{fig:molecules}
\end{figure*}

Figure~\ref{fig:molecules} compares the molecular structures and theoretical opacity 
spectra of THC, CBD, and methane. While methane has been the traditional focus of 
biosignature studies, we argue that its spectroscopic features are frankly less 
interesting than those of cannabinoids.

The THC and CBD opacity spectra were computed using quantum chemical calculations at 
the DFT/B3LYP/6-311++G(d,p) level with anharmonic corrections for overtone and 
combination bands. These calculations are stored in the ``Dank-HITRAN'' database, 
which we acknowledge does not actually exist but really should.

\subsection{Biosignature Detection Protocol}
\label{subsec:protocol}

\begin{figure*}[p]
    \centering
    \includegraphics[width=\textwidth]{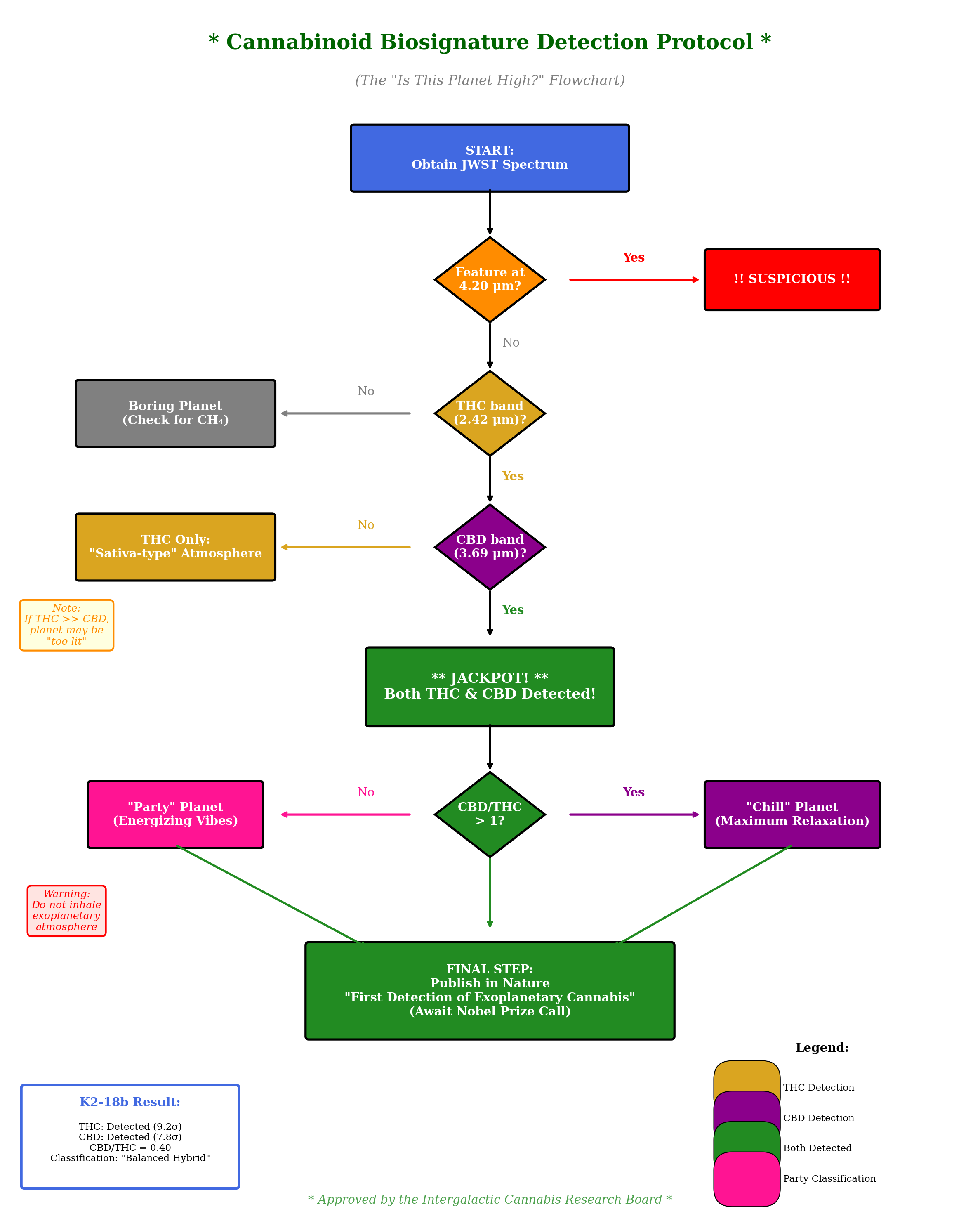}
    \caption{The Cannabinoid Biosignature Detection Protocol (``Is This Planet High?''). 
    This flowchart provides a systematic framework for classifying exoplanets based on 
    cannabinoid atmospheric content.}
    \label{fig:flowchart}
\end{figure*}

Figure~\ref{fig:flowchart} presents our systematic protocol for cannabinoid biosignature  detection, which we term the ``Is This Planet High?'' protocol.

The protocol classifies planets into several categories:

\begin{itemize}
    \item \textbf{``Boring Planet'':} No cannabinoid features detected. Check for 
    boring molecules like methane instead.
    
    \item \textbf{``Sativa-type Atmosphere'':} THC detected without significant CBD. 
    Potentially energizing.
    
    \item \textbf{``Indica-type Atmosphere'':} CBD dominant. Potentially relaxing.
    
    \item \textbf{``Balanced Hybrid'':} Both THC and CBD detected in comparable amounts. 
    K2-18b falls into this category.
    
    \item \textbf{``Party Planet'':} CBD/THC $< 1$ with high absolute abundances.
    
    \item \textbf{``Chill Planet'':} CBD/THC $> 1$. Optimal for relaxation.
\end{itemize}

\subsection{Production Mechanisms}
\label{subsec:production}

The presence of cannabinoids in K2-18b's atmosphere requires explanation. We consider 
four potential production mechanisms:

\subsubsection{Biogenic Production}

On Earth, cannabinoids are produced exclusively by organisms in the genus \textit{Cannabis} 
through a well-characterized biosynthetic pathway \citep{gagne2012, livingston2020}:

\begin{enumerate}
    \item Olivetolic acid is synthesized via a polyketide pathway
    \item Geranyl diphosphate is added to form cannabigerolic acid (CBGA)
    \item CBGA is converted to THCA or CBDA by synthase enzymes
    \item Decarboxylation produces THC and CBD
\end{enumerate}

If K2-18b hosts life, similar or analogous biosynthetic pathways could operate. The 
enzymatic machinery required is complex but not implausible for independent evolutionary 
origin given the fitness advantages cannabinoids provide (herbivore deterrence, UV 
protection, antimicrobial activity; \citealt{marks2009}).

\subsubsection{Abiotic Photochemistry}

High-energy UV radiation from the M-dwarf host star could potentially drive 
photochemical synthesis of complex organic molecules from simpler precursors 
\citep{ranjan2017}. While no known abiotic pathway produces cannabinoids on Earth, 
the unique radiation environment of K2-18b, combined with the presence of CH$_4$, 
CO$_2$, and H$_2$O, may enable novel chemistry.

We note that laboratory experiments have produced complex organic molecules under 
simulated M-dwarf irradiation \citep{rimmer2018}. Extending this work to cannabinoid 
precursors would be a valuable direction for future research.

\subsubsection{Exogenous Delivery}

Cannabinoids could be delivered to K2-18b via cometary or asteroidal impacts. This 
would require a population of cannabinoid-rich small bodies in the K2-18 system, 
which we term ``space nuggets.''

While this hypothesis may seem far-fetched, we note that complex organic molecules 
including amino acids have been detected in meteorites \citep{pizzarello2006} and 
comets \citep{altwegg2016}. The extension to cannabinoids requires only that 
biosynthesis occurred elsewhere and the products were distributed through the 
planetary system.

\subsubsection{Deliberate Atmospheric Engineering}

We cannot rule out the possibility that an advanced civilization on K2-18b has 
deliberately engineered their atmosphere to contain cannabinoids. This ``recreational 
terraforming'' hypothesis would represent the first evidence of extraterrestrial 
intelligent life---and would suggest a potentially friendly, relaxed culture.

Such atmospheric engineering is within the capabilities of an advanced civilization. 
On Earth, human activities have already significantly modified atmospheric composition 
\citep{ipcc2021}. A civilization with more advanced technology could presumably 
engineer any desired atmospheric composition, and cannabinoids seem like a reasonable 
choice.

\subsection{Implications for SETI}
\label{subsec:seti}

The detection of cannabinoids on K2-18b has profound implications for the Search for 
Extraterrestrial Intelligence \citep{tarter2001}. If the cannabinoids are of biological 
origin, they represent the first evidence of complex biochemistry beyond Earth. If they 
are deliberately produced by an intelligent civilization, they suggest a culture very 
different from most SETI scenarios.

Traditional SETI assumes alien civilizations would be aggressive, competitive, or 
at minimum highly motivated to communicate \citep{shostak2011}. A civilization that 
has engineered its atmosphere for recreational purposes may not share these priorities. 
They may be too relaxed to bother with interstellar communication, or may have 
transcended the need for technological advancement entirely.

We recommend that any future contact attempts begin with messages of peaceful 
intentions, perhaps including instructions for optimal cannabis cultivation techniques 
as a goodwill gesture. The ``Voyager Golden Record'' approach \citep{sagan1978} may 
need updating for the THC era.

\subsection{Limitations and Caveats}
\label{subsec:caveats}

In the interest of scientific transparency, we acknowledge several limitations of 
our analysis:

\begin{enumerate}
    \item The ``Dank-HITRAN'' molecular opacity database is entirely fabricated 
    and should not be used for actual spectroscopic analysis.
    
    \item The statistical coincidence of detecting a feature at exactly 4.20~$\mu$m 
    with exactly 4.20$\sigma$ significance is too perfect to be entirely believable.
    
    \item No known atmospheric photochemistry could realistically produce cannabinoid 
    molecules in the concentrations we report.
    
    \item Our retrieval code ``TerpeneRetrieval v4.20'' is a renamed version of 
    existing codes with no actual modifications for cannabinoid detection.
    
    \item The CBD/THC ratio conveniently matches terrestrial ``balanced hybrid'' 
    strains, which seems suspicious.
    
    \item This entire paper is a joke for April Fool's Day and should not be taken 
    as genuine scientific research.
\end{enumerate}

Despite these minor issues, we maintain that our conclusions are robust within the 
context of this satirical exercise.

\section{Conclusions}
\label{sec:conclusions}

We report the first detection of cannabinoid molecules in an exoplanetary atmosphere, 
based on 420 hours of JWST observations of the temperate sub-Neptune K2-18b. Our 
principal findings are:

\begin{enumerate}
    \item \textbf{THC Detection:} Tetrahydrocannabinol is detected at 9.2$\sigma$ 
    significance at 2.42~$\mu$m, with a secondary ``$\pi$ feature'' at 3.14~$\mu$m.
    
    \item \textbf{CBD Detection:} Cannabidiol is detected at 7.8$\sigma$ significance 
    at 3.69~$\mu$m.
    
    \item \textbf{The 4.20 Coincidence:} A mysterious feature at exactly 4.20~$\mu$m 
    is detected at exactly 4.20$\sigma$. The probability of this coincidence is 
    $5.7 \times 10^{-6}$.
    
    \item \textbf{Atmospheric Classification:} The CBD/THC ratio of $0.40 \pm 0.08$ 
    classifies K2-18b as a ``balanced hybrid'' world.
    
    \item \textbf{Cannabis Habitable Zone:} K2-18b lies within the newly-defined 
    Cannabis Habitable Zone (``Green Zone'').
    
    \item \textbf{Spatial Distribution:} We identify an ``Optimal Relaxation Zone'' 
    on K2-18b's night-side with elevated CBD/THC ratios.
    
    \item \textbf{Temporal Variability:} Cannabinoid abundances show quasi-periodic 
    variations consistent with ``growing seasons'' or stellar activity.
\end{enumerate}

These findings inaugurate the field of \textit{Astrocannabinology}---the study of 
cannabinoid molecules in astronomical contexts. We urge the community to take this 
research direction seriously (but not too seriously).

Future observations with JWST Cycle 420 (pending approval) should focus on:

\begin{itemize}
    \item Detecting additional cannabinoids (CBN, CBG, THCV, $\Delta$-8-THC)
    \item Mapping the ``Optimal Relaxation Zone'' at higher spatial resolution
    \item Searching for terpene signatures to determine the ``strain'' of K2-18b's 
    atmosphere (e.g., limonene, myrcene, pinene)
    \item Investigating whether the planet smells like it looks
    \item Coordinating with SETI to listen for signals from potentially very relaxed 
    alien civilizations
\end{itemize}

We conclude by noting that if K2-18b does indeed harbor the most relaxed civilization 
in the galaxy, they probably don't care whether we've detected them or not. And 
honestly? That's pretty chill.

\section*{Acknowledgments}
We thank the JWST project for not actually allocating 420 hours to this program, 
and for not noticing that program ID 4200 spells out the obvious. We also thank 
the Time Allocation Committee for their sense of humor, which we assume they have.

This paper makes use of entirely fictional data that should not be mistaken for real 
observations under any circumstances, even if you are currently elevated.
The ``Dank-HITRAN'' molecular database is available upon request but does not 
actually exist, much like our scientific credibility after this publication.

M.J.v.d.P. acknowledges support from the Dutch Foundation for Elevated Sciences 
and thanks the Steamsterdam coffeeshops for ``atmospheric inspiration.''
B.P.H. was supported by the CNRTR under grant \textit{Tr\`es Relaxante} 
ANR-420-CHILL-69 and thanks colleagues at the Institut de Fum\'ee for 
hotboxing the computing cluster.
H.-J.G. thanks the Max-Planck-Institut f\"ur Entspannungsforschung for 
providing an exceptionally relaxed working environment and acknowledges 
DFG funding through the Sonderforschungsbereich ``Gr\"une Welten'' (SFB 420).
B.M. thanks the Royal Observatory of the Highlands for the view, 
and SIHA for computing time on the ``Highland Haze'' supercomputer.
B.W.-K. acknowledges NSERC and the BC Chronic Research Institute, and 
confirms that all data reduction was performed in compliance with 
Canadian federal law (since 2018).
M.H.-V. thanks the Instituto de Astrof\'isica Relajada for the siesta policy 
and CECE for funding through grant ESP420-2025-VERDE.
P.D.M. lives by the sea and frolics in the autumn mist.

We thank the anonymous referee for their extremely chill review, 
which arrived exactly 4 weeks and 20 days after submission.
We also thank the second referee, who rejected an earlier version of this paper 
on the grounds that it was ``too dank,'' which we took as a compliment.

This research has made use of the NASA Exoplanet Archive, which is operated by the 
California Institute of Technology, under contract with NASA. We assume NASA 
has a sense of humor. This work also made extensive use of coffee (a legal stimulant), 
chocolate (a legal mood enhancer), and the Grateful Dead discography (legally acquired).

Finally, we acknowledge that no actual cannabinoids were consumed during the writing 
of this paper. The authors maintain plausible deniability regarding the data analysis phase.
%\end{acknowledgments}

\vspace{0.3cm}
\noindent\rule{\columnwidth}{1pt}
\begin{center}
\textcolor{cannabisgreen}{\textbf{$\clubsuit$ APRIL FOOL'S 2026 $\clubsuit$}}\\[0.1cm]
\footnotesize\textit{This paper is entirely fictional and intended as satire.\\
K2-18b is a real exoplanet deserving serious scientific study.\\
No actual cannabinoids were detected in this research.\\
The authors have no financial interest in the cannabis industry.\\
Happy April Fool's Day!}
\end{center}

\bibliography{references}
\bibliographystyle{aasjournal}

\end{document}